# Synthesis, Infra-red, Raman, NMR and structural characterization by X-ray Diffraction of $[C_{12}H_{17}N_2]_2CdCl_4$ and $[C_6H_{10}N_2]_2Cd_3Cl_{10}$ compounds

Iskandar Chaabane*, Faouzi Hlel and Kamel Guidara

Address: Laboratoire de l'état solide, Faculté des Sciences de Sfax, B. P. 802, 3018 Sfax, Tunisia

Email: Iskandar Chaabane* - chaabane.iskandar@yahoo.com; Faouzi Hlel - faouzihlel@yahoo.fr;
Kamel Guidara - kamelguidara@yahoo.fr

* Corresponding author

## Abstract

The synthesis, infra-red, Raman and NMR spectra and crystal structure of 2, 4, 4-trimethyl-4, 5-dihydro-3H-benzo [b] [1,4] diazepin-1-ium tetrachlorocadmate, $[C_{12}H_{17}N_2]_2CdCl_4$ and benzene-1,2-diaminium decachlorotricadmate(II) $[C_6H_{10}N_2]_2Cd_3Cl_{10}$ are reported.

The $[C_{12}H_{17}N_2]_2CdCl_4$ compound crystallizes in the triclinic system ($P\bar{1}$ space group) with Z = 2 and the following unit cell dimensions: a = 9.6653(8) Å, b = 9.9081(9) Å, c = 15.3737(2) Å, $\alpha$ = 79.486(1)°, $\beta$ = 88.610(8)° and $\gamma$ = 77.550(7)°. The structure was solved by using 4439 independent reflections down to R value of 0.029. In crystal structure, the tetrachlorocadmiate anion is connected to two organic cations through N-H...Cl hydrogen bonds and van der Waals interaction as to build cation-anion-cation cohesion. The $[C_6H_{10}N_2]_2Cd_3Cl_{10}$ crystallizes in the triclinic system ($P\bar{1}$ space group). The unit cell dimensions are a = 6.826 (5)Å, b = 9.861 (7)Å, c = 10.344 (3)Å, $\alpha$ = 103.50 (1)°, $\beta$ = 96.34 (4)° and $\gamma$ = 109.45 (3)°, Z = 2. The final R value is 0.053 (Rw = 0.128). Its crystal structure consists of organic cations and polymeric chains of $[Cd_3Cl_{10}]^{4-}$ anions running along the [011] direction, in the $[C_6H_{10}N_2]_2Cd_3Cl_{10}$ compounds hydrogen bond interactions between the inorganic chains and the organic cations, contribute to the crystal packing.

## 1. Introduction

The synthesis of low-dimensional mixed inorganic-organic materials enables both the inorganic and the organic components on the molecular scale to be optimised and thus to exhibit specific properties, such as electronic, optical, thermal and catalytic [1,2]. Among halometallates (II), chlorocadmates (II) have been of special interest for their structural flexibility. They can occur as simple tetrahedral anions $CdX_4^{2-}$ or form the backbone of chain polymers. This is due to the fact that the $Cd^{2+}$ ion, being a $d^{10}$, exhibits a great variety of coordination numbers and geometries, depending on crystal packing and hydrogen bonding, as well as halide dimensions [3-8]. Although two-dimensional polymeric chlorocadmates (II) have been widely studied, much less is known about one dimensional linear chain compounds. The major difficulty concerns the fact that the structural possibilities of bidimensional chlorocadmates(II) are somewhat restricted with respect to monodimensional ones, which present a panoply of different structural arrangements. A wide variety of stoichiometries belong to this class of compounds, including $[CdCl_3]$, $[Cd_3Cl_8]$, $[Cd_2Cl_5]$, $[Cd_2Cl_6]$, $[Cd_2Cl_7]$, $[Cd_3Cl_7]$, $[Cd_3Cl_{10}]$, $[Cd_5Cl_{14}]$, $[CdCl_3(OH_2)]$, $[Cd_2Cl_5(OH_2)]$, $[Cd_2Cl_6(OH_2)_2]$, $[Cd_2Cl_7(OH_2)]$ and $[Cd_5Cl_{12}(OH_2)_2]$ which makes the crystal chemistry of chlorocadmates (II) extremely diverse and complex. Common features of all these compounds are the invariable presence of octahedral $(CdCl_6)$ and/or $[CdCl_5(OH_2)]$ units in the anhydrous and hydrated species, respectively. Ionized in some different ways, such as sharing triangular faces, edges or vertexes, giving rise to infinite chains usually running parallel along a crystallographic axis. A network of hydrogen bonding, involving organic cations generally connects the columnar stacks of chains together and stabilizes the whole crystal structure [9-16].

In the present paper we report on the synthesis, the structural and spectroscopic characterizations of the both $[C_6H_{10}N_2]_2Cd_3Cl_{10}$ and $[C_{12}H_{17}N_2]_2CdCl_4$ compounds.

## 2. Experimental
### *2.1. Preparation of [C$_{12}$H$_{17}$N$_2$]$_2$CdCl$_4$ and [C$_6$H$_{10}$N$_2$]$_2$Cd$_3$Cl$_{10}$ compound*

$[C_{12}H_{17}N_2]_2CdCl_4$ sample (denoted 1) was prepared by mixing the organic compound 1,2-phenylenediamine, dissolved in acetone, with $CdCl_2$, dissolved in hydrochloric acid solution (1 M), in molar ratio 2:1. By slow evaporation at room temperature, yellow crystals suitable for X-ray single crystal analysis were obtained.

Schematically the reaction proposed in order to justify obtaining $[C_6H_{10}N_2]_2Cd_3Cl_{10}$ sample is shown in figure 1.

$[C_6H_{10}N_2]_2Cd_3Cl_{10}$ (denoted 2) sample was prepared by mixing $CdCl_2$, dissolved in hydrochloric acid solution (1 M), and the organic compound 1,2-phenylenediamine, in molar ratio 2:1. By slow evaporation at room temperature, red crystals suitable for X-ray single crystal analysis were obtained.

Schematically the reaction proposed in order to justify obtaining $[C_6H_{10}N_2]_2Cd_3Cl_{10}$ sample is shown in figure 2.

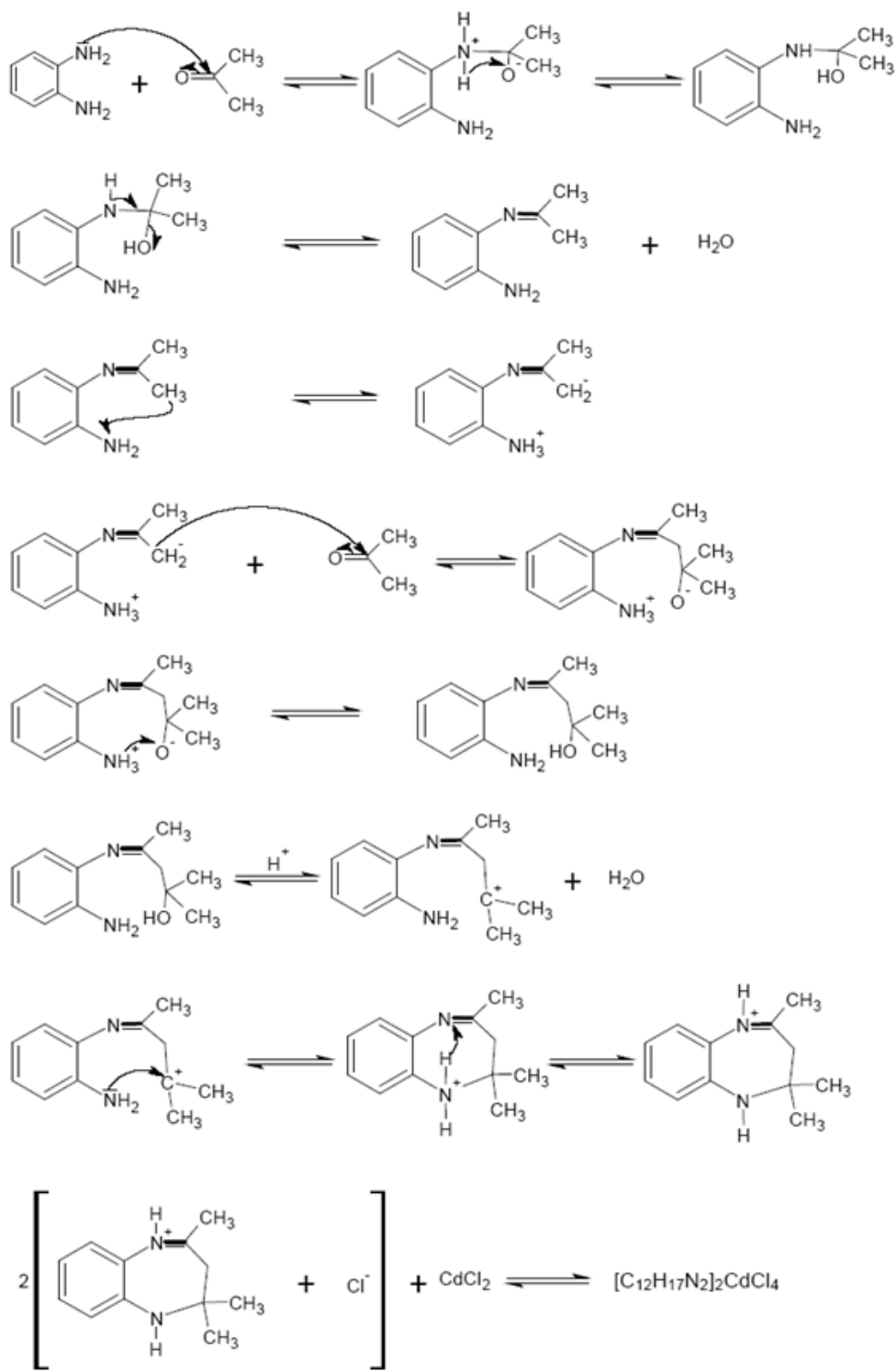

**Figure 1**
Schematically the reaction proposed in order to justify obtaining [C$_{12}$H$_{17}$N$_2$]$_2$CdCl$_4$ sample.

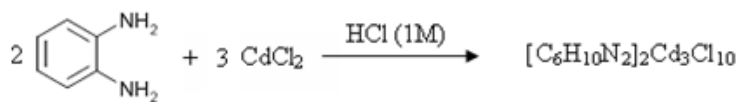

**Figure 2**
Schematically the reaction proposed in order to justify obtaining [$C_6H_{10}N_2$]$_2$Cd$_3$Cl$_{10}$ sample.

## 2.2. Analysis

The infra-red spectrum was recorded in the 400–4000 cm$^{-1}$ range with a Perkin-Elmer FT-IR 1000 spectrometer using samples pressed in spectroscopically pure KBr pellets.

The Raman spectrum of [$C_{12}H_{17}N_2$]$_2$CdCl$_4$ and [$C_6H_{10}N_2$]$_2$Cd$_3$Cl$_{10}$ samples were recorded respectively on a Kaiser Optical System spectrometer model-Hololab 5000R in the region 70–3500 cm$^{-1}$ and HR 800 spectrometer in the region 150–2000 cm$^{-1}$.

For both compounds $^{111}$Cd CP-MAS NMR spectra were measured on powdered sample at 63.648 MHz (7.1 T) with Bruker WB 300 MAS FT-NMR spectrometer. A single pulse sequence was used for all the measurements. The acquisition parameters for compound 1 and 2 were as follows; a 10.5 µs pulse length, 5.0 s pulse delays and 512 and 1024 scan per spectrum respectively. Samples in cylindrical zirconia rotors were spun at spinning rates at 4 KHz and 11 KHz respectively. Chemical shifts were referenced to Cd(ClO$_4$)$_2$ aqueous solution at 0 ppm.

## 2.3. Crystallographic studies

For compound 1 (0.58 mm × 0.38 mm × 0.25 mm) and compound 2 (0.57 × 0.4 × 0.3 mm$^3$) prismatic crystals were selected by optical examination and mounted on an Enraf-Nonius CAD4 four-circle diffractometer. For both compound 1 and 2 the unit-cell parameters were determined from automatic centering of 25 reflections (12 < θ < 15° and 11 < θ < 16° respectively) and refined by least-squares method. Intensities were collected with graphite monochromatized Mo K$_α$ radiation, using ω/2θ scan-technique.

Two reflections were measured every 2 hrs as orientation and intensity control. No absorption corrections were made for compound 1 and an empirical absorption correction was applied for compound 2 using a method based upon ABSDIF data (transmission range of 0.1611–0.4206). The both structures were solved by Patterson methods, using SHELXS-86 [17] in the space group $P\bar{1}$. All the refinement calculations were performed with the SHELXL-93 [18] computer program. Cd and Cl atoms were first located. The atomic positions of nitrogen, carbon and hydrogen of the organic groups were subsequently found by difference Fourier syntheses. Details of the crystal structure analysis are reported in Table 1.

Table 1: Summary of crystal data, intensity measurements and refined parameters for [$C_{12}H_{17}N_2$]$_2$CdCl$_4$ and [$C_6H_{10}N_2$]$_2$[Cd$_3$Cl$_{10}$] compounds.

| Crystal data | Compound 1 | Compound 2 |
| --- | --- | --- |
| Formula Formula weight (g.mol$^{-1}$) | [$C_{12}H_{17}N_2$]$_2$CdCl$_4$ 632.75 | [$C_6N_2H_{10}$]$_2$Cd$_3$Cl$_{10}$ 455 |
| Color/shape | yellow/parallelepiped | red/parallelepiped |
| Crystal dimensions (mm$^3$) | 0.58 × 0.38 × 0.25 | 0.57 × 0.4 × 0.3 |
| Crystal system | triclinic | triclinic |
| Space group | $P\bar{1}$ | $P\bar{1}$ |
| Cell parameters from 25 | 12 < θ(°) < 15 | 11 < θ° < 16 |
|  | a = 9.687 (8) Å, | a = 6.826 (5)Å, |
|  | b = 9.912 (9) Å, | b = 9.861 (7)Å |
|  | c = 15.40 (2) Å | c = 10.344 (3)Å |
|  | α = 79.4 (1)°, | α = 103.50 (1)° |
|  | β = 88.73 (8)° | β = 96.34 (4)° |
|  | γ = 77.82 (7)° | γ = 109.45 (3)° |
|  | V = 1420 Å$^3$ | V = 624.8 (8) Å$^3$ |
|  | Z = 2, μ = 1.164 mm$^{-1}$ | Z = 2, μ = 3.609 mm$^{-1}$ |
| **Intensity measurements** |  |  |
| Temperature (K) | 293(2) | 293(2) |
| Radiation, λ (Å), monochromator | MoK$_α$, 0.71069, graphite plate | MoK$_α$, 0.71069, graphite plate |
| Scan angle (°) | 0.8 + 0.35 tg(θ) | 0.8 + 0.35 tg(θ) |
| 2θ range (°) | 1.5 – 27 | 2 – 27 |
| Range of h, k, l | -12→12, -2→12, 0→19 | -8→8, -12→12, -13→13 |
| Standards reflections | (6 4 0) and (2 6 0) | (-1 6 1) and (-1 5 1) |
| Frequency | 60 mn |  |
| Reflections collected/unique | 4541/4439 (Rint = 0.0219) | 5436/2718 (Rint = 0.0019) |
| **Structure determination** |  |  |
| Absorption correction | wasn't applied | ABSDIF T$_{min/max}$: 0.1611/0.4206 |
| Structure resolution | Patterson methods SHELXS86 | Patterson methods SHELXS86 |
| Structure refinement with | SHELXL-97 | SHELXL-97 |
| Observed reflections [Fo > 2σ(Fo)] | 3823 | 2371 |
| Refinement | F$^2$ full matrix | F$^2$ full matrix |
| Refined parameters | 435 | 174 |
| Goodness of fit | 1.035 | 1.565 |
| Final R and Rw | 0.029, 0.081 | 0.053, 0.128 |
| Final R and Rw for all data | 0.0378, 0.0864 | 0.0615, 0.1315 |
| Largest feature diff. map | 0.668, -0.453 e Å$^{-3}$ | 3.256, -2.636 e Å$^{-3}$ |

For compound 1: w = 1/[σ$^2$ (Fo)$^2$ + (0.0315 P)$^2$ + 0 P] where P = [Fo$^2$ + 2 Fc$^2$]/3
For compound 2: w = 1/[σ$^2$ (Fo)$^2$+(0.0507P)$^2$ + 0 P] where P = [Fo$^2$ + 2 Fc$^2$]/3

## 3. Results and discussions

### 3.1. Infra-red and Raman spectroscopy

#### 3.1.1 [$C_{12}H_{17}N_2$]$_2$CdCl$_4$ (1)

Figures 3a and 3b show IR and Raman spectra respectively of the reported compound at room temperature. A detailed assignment of all the bands is difficult, but we can attribute some of them by comparison with similar compounds [19-21]. The assignments of the bands observed in the infrared and Raman spectra of [$C_{12}H_{17}N_2$]$_2$CdCl$_4$ are listed in Table 2.

The principal bands are assigned to the internal modes of organic cation. The C = C bands exhibit torsion vibration at 457 cm$^{-1}$ in IR and 460 cm$^{-1}$ in Raman, and stretching vibration at 1580 cm$^{-1}$ in IR. The bands observed at 763, 955 and 765, 953 cm$^{-1}$ in IR and Raman respectively are ascribed to CH wagging mode. Those observed at 1166, 1292, 1461 cm$^{-1}$ and 1175, 1288,

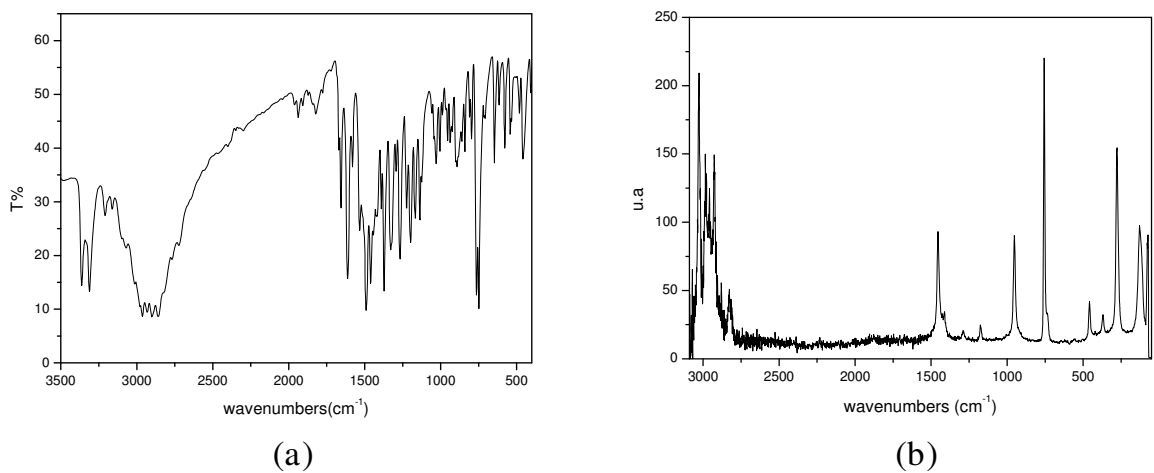

**Figure 3**
**a:** Infrared spectrum of $[C_{12}H_{17}N_2]_2CdCl_4$. **b:** Raman spectrum of $[C_{12}H_{17}N_2]_2CdCl_4$.

1454 cm$^{-1}$ in IR and Raman respectively are ascribed to CH bending vibration. The CH stretching vibration are observed at 2931, 2963, 2978 cm$^{-1}$ in IR and 2926, 2957, 2985 cm$^{-1}$ in Raman. The bands observed at 3160 and 3208 cm$^{-1}$ in IR are associated to the asymmetric NH stretching out of plane, those observed at 3313 and 3363 cm$^{-1}$ in IR are ascribed to symmetric NH stretching in plane.

Table 2: Infrared and Raman spectral data (cm$^{-1}$) and band assignments for $[C_{12}H_{17}N_2]_2CdCl_4$.

| IR wavenumbers (cm$^{-1}$) | Raman wavenumbers (cm$^{-1}$) | Assignment |
|---|---|---|
|  | 74 | $(CdCl_4^{2-})$ Bend. |
|  | 78 | $(CdCl_4^{2-})$ Bend. |
|  | 117 | Rotation $(C_{12}H_{17}N_2^+)$ |
|  | 128 | Rotation $(C_{12}H_{17}N_2^+)$ |
|  | 279 | $CdCl_4^{2-}$ Symmetric stretch |
| 457 | 460 | C = C torsion |
| 713 | 738 | CC torsion |
| 763 | 756 | CH and NH Wagg. |
| 955 | 953 | CH Wagg. |
| 1166 | 1175 | CH and $CH_3$ Bend. |
| 1292 | 1288 | CH and NH Bend., CN Str. |
| 1461 | 1454 | CC and CN Str., CH Bend. |
| 1580 |  | C = C Str. |
| 2931 | 2926 | CH Str. |
| 2963 | 2957 | CH Str. |
| 2978 | 2985 | CH Str. |
|  | 3026 | CH Str. (+) |
|  | 3030 | CH Str. (+) |
| 3160 |  | Asymmetric NH Str. (-) |
| 3208 |  | Asymmetric NH Str. (-) |
| 3313 |  | Symmetic NH Str. (+) |
| 3363 |  | Symmetic NH Str. (+) |

Bend., bending; Str., stretching; Wagg., wagging; (-) out-of-plane; (+) in plane.

The bands corresponding to the internal vibrational modes of the (CdCl$_4$) anions: $\nu_1$, $\nu_2$, $\nu_3$ and $\nu_4$ appear in the Raman spectral region below 300 cm$^{-1}$. The intense band observed at 279 cm$^{-1}$ is assigned to the $\nu_1$ mode; while the band observed at 78 cm$^{-1}$ is assigned to the $\nu_4$ mode. Finally, the band appearing at 74 cm$^{-1}$ is assigned to the $\nu_2$ mode.

The infrared and Raman study confirms the presence of the organic group $C_{12}H_{17}N_2$ and the tetrahedral anion $CdCl_4^{2-}$.

### 3.1.2 [C$_6$H$_{10}$N$_2$]$_2$Cd$_3$Cl$_{10}$ (2)

FTIR and Raman spectra of [C$_6$H$_{10}$N$_2$]$_2$Cd$_3$Cl$_{10}$ (2) have been recorded (see Figure 4a and 4b) at room temperature. A detailed assignment of all the bands is difficult but we can attribute some of them by comparison with similar compounds [19].

In IR spectrum, the band observed at ~3500 cm$^{-1}$ can be attributed to symmetric and asymmetric NH stretching vibrations. The characteristic ν(CH) modes of the aromatic ring are observed as expected in the 3170-2680 cm$^{-1}$ spectral regions. The bands observed between ~1540 and ~1620 cm$^{-1}$ are ascribed to C = C stretching mode. The intense band observed at 1490 cm$^{-1}$ is associated with the vibrations of the C-N-H group mixed with C-C stretching and CH bending. The intense band observed at 1470 cm$^{-1}$ is attributed to C-C and C-N stretching and CH bending vibrations. A very strong band observed at 1310 cm$^{-1}$ results purely from interaction between the N-H and C-H bending and C-N stretching vibrations. The band observed between 1130 cm$^{-1}$ and 1240 cm$^{-1}$ is attributed to CH bending vibrations. The symmetric out of plane CH vibration (CH-wagging) creates a very strong band at 772 cm$^{-1}$ whereas weak bands observed at 861 cm$^{-1}$ and 876 cm$^{-1}$ were attributed to CCC bending and C-N stretching. The medium bands observed in the region between 519 and 581 cm$^{-1}$ are commonly attributed to ring torsion. The strong band observed at 443 cm$^{-1}$ is ascribed to C = C torsion vibration. These vibrational absorptions are listed in Table 3.

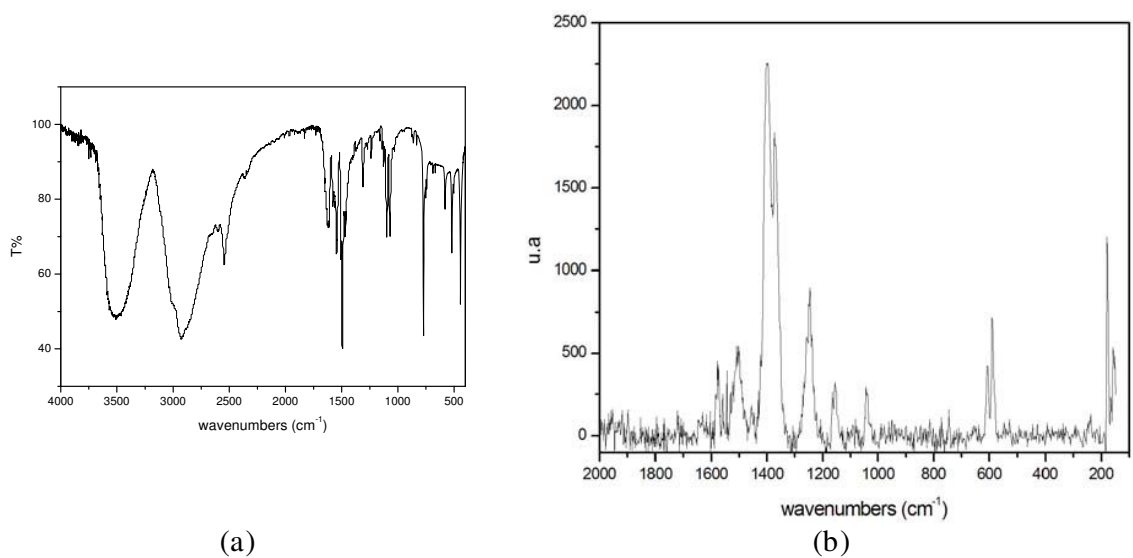

**Figure 4**
**a:** Infrared spectrum of [C$_6$H$_{10}$N$_2$]$_2$[Cd$_3$Cl$_{10}$]. **b:** Raman spectrum of [C$_6$H$_{10}$N$_2$]$_2$[Cd$_3$Cl$_{10}$].

The bands corresponding to the internal vibrational modes of the (CdCl$_6$) anions appear in the Raman spectral region below 300 cm$^{-1}$. The intense band observed at 158 cm$^{-1}$ is assigned to the Cd-Cl bending; while the band observed at 180 cm$^{-1}$ is assigned to the Cd-Cl (equatorial) bending [22]. These vibrational absorptions are given in Table 3.

The infrared and Raman study confirms the presence of the organic group C$_6$H$_{10}$N$_2$ and the octahedral anion CdCl$_6$.

Table 3: Infrared and Raman spectral data (cm$^{-1}$) and band assignments for [C$_6$H$_{10}$N$_2$]$_2$[Cd$_3$Cl$_{10}$] sample.

| IR wavenumbers (cm$^{-1}$) | Raman wavenumbers (cm$^{-1}$) | Assignment |
|---|---|---|
|  | 158 | Cd-Cl Bend. |
|  | 180 | Cd-Cl (equatorial) Bend. |
| 443 |  | C = C torsion |
| 503 |  |  |
| 519 | 592 | aromatic ring torsion |
| 581 | 606 |  |
| 668 |  | CCC Bend. |
| 688 |  |  |
| 743 |  | CC torsion |
| 752 |  |  |
| 772 |  | CH and NH Wagg. |
| 833 |  | CH Wagg. |
| 861 |  | NH Wagg. |
| 876 |  | CH Wagg. |
| 1070 | 1044 | CC Str., CH and CCC Bend. |
| 1100 |  |  |
| 1130 | 1155 | CH Bend. |
| 1140 | 1165 |  |
| 1240 | 1247 |  |
| 1310 | 1373 | CC and CN Str. |
| 1470 | 1400 |  |
| 1490 |  | CH Bend., CC and CN Str. |
| 1540 | 1505 | C = C Str. |
| 1580 | 1545 |  |
| 1620 | 1575 |  |
| 2920 |  | CH Str. |
| 3500 |  | Asymmetric and Symmetic NH Str. |

Bend., bending; Str., stretching; Wagg., wagging;

### 3.2. NMR measurements

The [111]Cd MAS NMR spectra of compound 1 and 2 are shown respectively in Figures 5a and 5b. The spectrum of compound 1 is composed of one broad peak and two spinning side bands. The $\delta_{iso}$ value is equal to 468.01 ppm indicating that $CdCl_4^{2-}$ tetrahedral is present in the sample [23]. Based on the [111]Cd MAS NMR measurements of several chlorocadmate crystals with known structures, Sakida et al. have shown that the $\delta_{iso}$ value for the $CdCl_6$ octahedra is equal to 183 ppm [24]. Hence, it may be concluded that the compound 2 is composed of $CdCl_6$ octahedra alone. The $\delta_{iso}$ value of $Cd(1)Cl_6$ octahedra is much smaller than that of $Cd(2)Cl_6$. This fact indicates that the octahedral anion-co-ordination around $Cd^{2+}$ has higher symmetry in the $Cd(1)Cl_6$ octahedra than in the $Cd(2)Cl_6$.

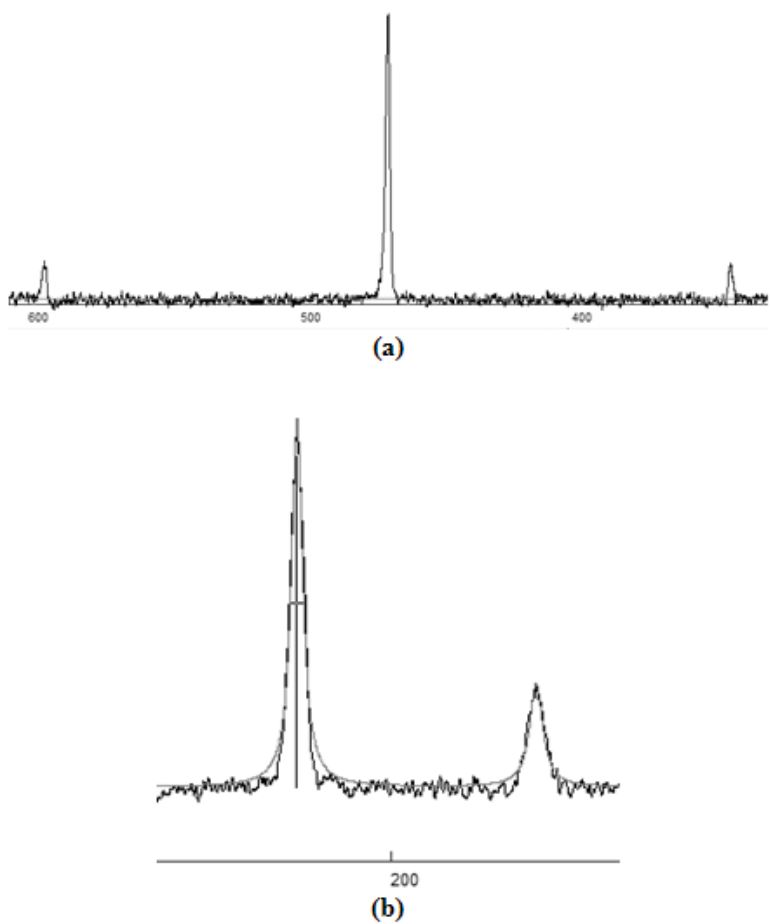

**Figure 5**
a: Cross polarization [111]Cd MAS NMR spectra of $[C_{12}H_{17}N_2]_2CdCl_4$ compound. b: Cross polarization [111]Cd MAS NMR spectra of $[C_6H_{10}N_2]_2[Cd_3Cl_{10}]$ compound.

### 3.3. Description of the structures

#### 3.3.1 [C$_{12}$H$_{17}$N$_2$]$_2$CdCl$_4$ (1)

Selected inter-atomic distances and angles are given in Tables 4, 5, 6 and 7, respectively.

The structure can be described by an alternation of organic and inorganic layers stacked in the c direction. The anionic layer is built up of tetrahedra of tetrachlorocadmate CdCl4$^{2-}$ sandwiched between two different organic layers. The first one located at z = 0 is formed by C12N2H17+ (a) cations. The second is build up of C12N2H17+ (b) cations observed at z = 1/2 (Figure 6). The benzene rings in cations one and two are not parallel. The angle between them is equal to 77.3 (1)°.

The material cohesion is assured by two different interactions:

- The bonding between the organic and inorganic layers is established by three different hydrogen bonds (Cl1...HN21-N21, Cl2...HN12-N12 and Cl2...HN11-N11). The N...Cl distances vary between 3.134 (4) Å and 3.691 (4) Å. We deduce that the hydrogen links are weak (Table 8) [25].

- The cohesion of the organic layer is obtained by van der Waals interaction between aromatic π-stacking. The separation between the planes of two aromatic cations from adjacent dimeric unit is 4.29 (6) Å [4,26] (Figure 6).

#### 3.3.1.1 Geometry and coordination of the tetrachlorocadmiate anion

CdCl$_4$$^{2-}$ tetrahedron presents a C$_1$ punctual symmetry. The geometrical features of CdCl$_4$ entity are comparable to those found in the Cambridge Structural Database (CSD) for other Cd(II) salts containing isolated [CdCl$_4$]$^{2-}$ tetrahedral anions [27]. The range of average Cd-Cl distances is 2.45–2.47 Å. In our case the Cd-Cl distances vary between 2.416(2) and 2.501(2) Å with a mean of 2.459(3) Å (Table 4). The Cl-Cd-Cl angle values are in the 101.81(9)° – 118.13(9)° range with a mean of 109.26(9)° (Table 6). Taking into account these parameters and considering the cal-

Table 4: Main inter-atomic distances (Å) in anionic part of [C$_{12}$H$_{17}$N$_2$]$_2$CdCl$_4$ and [C$_6$H$_{10}$N$_2$]$_2$[Cd$_3$Cl$_{10}$] compounds. (Esd are given in parentheses).

| Compound (1) | | Compound (2) | | | |
|---|---|---|---|---|---|
| Distance (Å) of CdCl$_4$ | | Distance (Å) of Cd(1)Cl$_6$ | | Distance (Å) of Cd(2)Cl$_6$ | |
| Cd-Cl(1) | 2.497(3) | Cd1–Cl3 | 2.551(1) | Cd2–Cl6 | 2.519(1) |
| Cd-Cl(2) | 2.501(2) | Cd1–Cl3[i] | 2.551(1) | Cd2–Cl7[i] | 2.583(1) |
| Cd-Cl(3) | 2.423(3) | Cd1–Cl5 | 2.669(1) | Cd2–Cl3 | 2.600(1) |
| Cd-Cl(4) | 2.416(2) | Cd1–Cl5[i] | 2.669(1) | Cd2-Cl4 | 2.694(1) |
| | | Cd1-Cl4[ii] | 2.689(1) | Cd2-Cl5[iv] | 2.710(1) |
| | | Cd1-Cl4[iv] | 2.689(1) | Cd2-Cl4[iii] | 2.739(1) |





Table 5: Main inter-atomic distances (Å) in cationic part of [C$_{12}$H$_{17}$N$_2$]$_2$CdCl$_4$ and [C$_6$H$_{10}$N$_2$]$_2$[Cd$_3$Cl$_{10}$] compounds. (Esd are given in parentheses).

| Compound (1) | | | | Compound (2) | |
|---|---|---|---|---|---|
| **Distance (Å) of C$_{12}$H$_{17}$N$_2^+$(a)** | | **Distance (Å) of C$_{12}$H$_{17}$N$_2^+$(b)** | | **Distance (Å) of C$_6$N$_2$H$_{10}$** | |
| C(11)-C(12) | 1.388(6) | C(21)-C(22) | 1.383(5) | C1-C6$^{vi}$ | 1.387(6) |
| C(12)-C(13) | 1.383(7) | C(22)-C(23) | 1.378(5) | C2-C1$^v$ | 1.367(7) |
| C(13)-C(14) | 1.373(8) | C(23)-C(24) | 1.396(6) | C2$^{vi}$-C5 | 1.381(8) |
| C(14)-C(15) | 1.356(7) | C(24)-C(25) | 1.352(6) | C3-C4 | 1.367(8) |
| C(15)-C(16) | 1.410(5) | C(25)-C(26) | 1.426(4) | C3-C6 | 1.395(7) |
| C(16)-C(11) | 1.406(4) | C(21)-C(26) | 1.415(5) | C4-C5$^v$ | 1.378(9) |
| C(11)-N(11) | 1.405(5) | C(21)-N(21) | 1.430(4) | C1-N1 | 1.473(6) |
| N(11)-C(17) | 1.482(4) | N(21)-C(27) | 1.291(5) | C6-N2 | 1.460(6) |
| C(17)-C(171) | 1.533(6) | C(27)-C(271) | 1.502(5) | | |
| C(17)-C(172) | 1.511(7) | C(27)-C(28) | 1.466(5) | | |
| C(17)-C(18) | 1.542(5) | C(28)-C(29) | 1.551(4) | | |
| C(18)-C(19) | 1.492(6) | C(29)-C(291) | 1.520(6) | | |
| C(19)-C(191) | 1.467(8) | C(29)-C(292) | 1.506(6) | | |
| C(19)-N(12) | 1.307(5) | C(29)-N(22) | 1.485(5) | | |
| C(16)-N(12) | 1.400(5) | N(22)-C(26) | 1.350(5) | | |

- *symmetry code*: i: -x+1,-y,-z+1; ii: x-1,y,z; iii: -x+2,-y,-z+1; vi: x+1,y,z;
v: -x+2,-y,-z+2; vi: x,y-1,z; vii: x,y+1,z;

culated average values of the Baur distortion indices [28] {ID Cd-Cl = 0.0157(1), ID Cl-Cd-Cl = 0.031(1) and ID Cl-Cl = 0.0157(1)}, we deduce that the CdCl$_4$ tetrahedron is slightly distorted.

*3.3.1.2 Geometry and coordination of the organic cations*

Ortep representations of C$_{12}$H$_{17}$N$_2^+$ (a) and C$_{12}$H$_{17}$N$_2^+$(b) cations showing ellipsoid thermal unrest are shown in Figures 7a and 7b, respectively.

Table 6: Main bond angles (°) in anionic part of [C$_{12}$H$_{17}$N$_2$]$_2$CdCl$_4$ and [C$_6$H$_{10}$N$_2$]$_2$[Cd$_3$Cl$_{10}$] compounds. (Esd are given in parentheses).

| Compound (1) | | Compound (2) | | | |
|---|---|---|---|---|---|
| **Angle (°) of CdCl$_4$** | | **Angle (°) of Cd(1)Cl$_6$** | | **Angle (°) of Cd(2)Cl$_6$** | |
| Cl(1)-Cd-Cl(2) | 101.81(9) | Cl3-Cd1-Cl3$^i$ | 180 | Cl6-Cd2-Cl7 | 92.55(4) |
| Cl(1)-Cd-Cl(3) | 108.85(9) | Cl3-Cd1-Cl5 | 90.46(4) | Cl6-Cd2-Cl3 | 96.22(4) |
| Cl(1)-Cd-Cl(4) | 110.14(9) | Cl3$^i$-Cd1-Cl5 | 89.54(4) | Cl7-Cd2-Cl3 | 96.02(4) |
| Cl(2)-Cd-Cl(3) | 109.78(9) | Cl3-Cd1-Cl5$^i$ | 89.54(4) | Cl6-Cd2-Cl4 | 94.15(4) |
| Cl(2)-Cd-Cl(4) | 106.87(8) | Cl3$^i$-Cd1-Cl5$^i$ | 90.46(4) | Cl7-Cd2-Cl4 | 166.97(4) |
| Cl(3)-Cd-Cl(4) | 118.13(9) | Cl5-Cd1-Cl5$^i$ | 180 | Cl3-Cd2-Cl4 | 94.35(4) |
| | | Cl3-Cd1-Cl4$^{ii}$ | 95.14(3) | Cl6-Cd2-Cl5$^{iv}$ | 92.92(4) |
| | | Cl3$^i$-Cd1-Cl4$^{ii}$ | 84.86(3) | Cl7-Cd2-Cl5$^{iv}$ | 85.10(4) |
| | | Cl5-Cd1-Cl4$^{ii}$ | 84.28(3) | Cl3-Cd2-Cl5$^{iv}$ | 170.72(4) |
| | | Cl5$^i$-Cd1-Cl4$^{ii}$ | 95.72(3) | Cl4-Cd2-Cl5$^{iv}$ | 83.41(3) |
| | | Cl3-Cd1-Cl4$^{iii}$ | 84.86(3) | Cl6-Cd2-Cl4$^{iii}$ | 178.10(4) |
| | | Cl3$^i$-Cd1-Cl4$^{iii}$ | 95.14(3) | Cl7-Cd2-Cl4$^{iii}$ | 89.22(4) |
| | | Cl5-Cd1-Cl4$^{iii}$ | 95.72(3) | Cl3-Cd2-Cl4$^{iii}$ | 82.93(4) |
| | | Cl5$^i$-Cd1-Cl4$^{iii}$ | 84.28(3) | Cl4-Cd2-Cl4$^{iii}$ | 84.23(4) |
| | | Cl4$^{ii}$-Cd1-Cl4$^{iii}$ | 180 | Cl5$^{iv}$-Cd2-Cl4$^{iii}$ | 87.88(3) |





Table 7: Main bond angles (°) in cationic groups of [C$_{12}$H$_{17}$N$_2$]$_2$CdCl$_4$ and [C$_6$H$_{10}$N$_2$]$_2$[Cd$_3$Cl$_{10}$] compounds. (Esd are given in parentheses).

| | Compound (1) | | | Compound (2) | |
| --- | --- | --- | --- | --- | --- |
| Angle (°) of C$_{12}$H$_{17}$N$_2$$^+$(a) | | Angle (°) of C$_{12}$H$_{17}$N$_2$$^+$(b) | | Angle (°) of C$_6$N$_2$H$_{10}$ | |
| C(13)-C(12)-C(11) | 121.9(4) | C(23)-C(22)-C(21) | 121.7(4) | C2$^v$-C1-C6$^{vi}$ | 120.8(5) |
| C(14)-C(13)-C(12) | 120.4(5) | C(22)-C(23)-C(24) | 118.8(4) | C1$^v$-C2-C5$^{vii}$ | 119.1(4) |
| C(15)-C(14)-C(13) | 119.6(4) | C(25)-C(24)-C(23) | 120.0(3) | C3-C4-C5$^v$ | 120.2(5) |
| C(14)-C(15)-C(16) | 120.9(4) | C(24)-C(25)-C(26) | 123.2(3) | C2$^{vi}$-C5-C4$^v$ | 120.9(5) |
| C(12)-C(11)-C(16) | 117.1(3) | C(21)-C(26)-C(25) | 115.4(3) | C1$^{vii}$-C6-C3 | 119.4(5) |
| N(11)-C(11)-C(16) | 120.4(4) | C(22)-C(21)-C(26) | 120.8(3) | C1$^{vii}$-C6-N2 | 121.7(4) |
| C(12)-C(11)-N(11) | 122.2(3) | C(26)-C(21)-N(21) | 124.7(3) | C3-C6-N2 | 118.8(4) |
| C(11)-N(11)-C(17) | 122.7(3) | C(22)-C(21)-N(21) | 114.4(3) | C2$^v$-C1-N1 | 118.5(4) |
| N(11)-C(17)-C(171) | 110.4(4) | C(27)-N(21)-C(21) | 129.3(4) | C6$^{vi}$-C1-N1 | 120.7(4) |
| N(11)-C(17)-C(172) | 108.3(3) | N(21)-C(27)-C(271) | 117.5(4) | | |
| C(172)-C(17)C(171) | 109.5(4) | N(21)-C(27)-C(28) | 120.1(3) | | |
| N(11)-C(17)-C(18) | 107.8(2) | C(28)-C(27)-C(271) | 122.3(3) | | |
| C(172)-C(17)-C(18) | 112.6(4) | C(27)-C(28)-C(29) | 115.1(3) | | |
| C(171)-C(17)-C(18) | 108.2(4) | C(291)-C(29)-C(28) | 109.7(3) | | |
| C(19)-C(18)-C(17) | 113.1(3) | C(292)-C(29)-C(28) | 109.1(3) | | |
| C(191)-C(19)-C(18) | 122.7(4) | C(292)-C(29)-C(291) | 111.0(3) | | |
| N(12)-C(19)-C(18) | 117.3(4) | N(22)-C(29)-C(28) | 108.9(3) | | |
| N(12)-C(19)-C(191) | 119.8(4) | N(22)-C(29)-C(291) | 113.1(4) | | |
| C(19)-N(12)-C(16) | 126.5(3) | N(22)-C(29)-C(292) | 104.9(3) | | |
| N(12)-C(16)-C(11) | 120.2(3) | C(26)-N(22)-C(29) | 128.6(3) | | |
| N(12)-C(16)-C(15) | 119.6(3) | N(22)-C(26)-C(21) | 129.3(3) | | |
| C(11)-C(16)-C(15) | 120.0(4) | N(22)-C(26)-C(25) | 115.2(3) | | |

- <u>symmetry code</u>: i: -x+1,-y,-z+1; ii: x-1,y,z; iii: -x+2,-y,-z+1; iv: x+1,y,z;
v: -x+2,-y,-z+2; vi: x,y-1,z; vii: x,y+1,z;

The C-C distances in the benzene of the first and second cations vary in the ranges 1.356(8)–1.411(2) Å and 1.352(6)–1.426(4) Å respectively (Table 5). The C-C-C angle values in the aromatic ring are in the range 117.1(3)° – 121.9(4)° and 115.4(3)° – 123.2(3)° for the first and second cation respectively (Table 7).

The benzene rings of those cations are distorted and slightly deviated from their medium plane.

The equation of the average plane for the first and second cations are, respectively, 7.11(1) x + 1.70(1) y - 9.96(2) z = 0.66 (3) and 4.55 (1) x - 1.99 (2) y + 11.80 (2) z = 4.91 (2). In the first and second cation, the average deviation of carbon atoms from the ideal aromatic ring is 0.0032 Å and 0.0064 Å, respectively. We deduce that the C$_{12}$H$_{17}$N$_2$$^+$(2) cation is more deformed.

### 3.3.2 [C$_6$H$_{10}$N$_2$]$_2$Cd$_3$Cl$_{10}$ (2)

Selected inter-atomic distances and angles are given in Tables 4, 5, 6 and 7, respectively.

The crystal structure of the [C$_6$H$_{10}$N$_2$]$_2$[Cd$_3$Cl$_{10}$] compound can be described by an alternation of organic and inorganic layers stacked in the [011] direction. The inorganic layer is built up

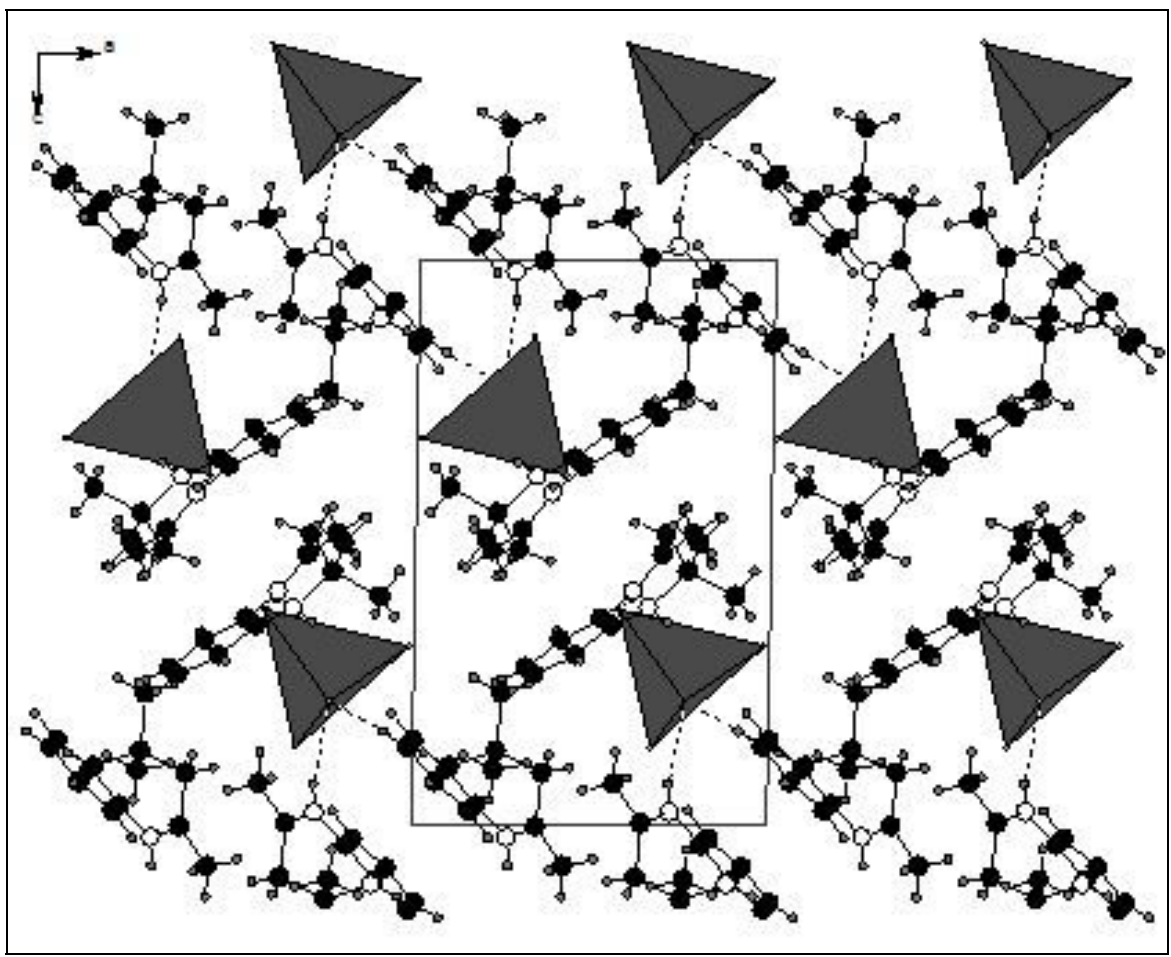

**Figure 6**
[010] view of the structure of $[C_{12}H_{17}N_2]_2CdCl_4$. The large empty circles represent nitrogen atoms, the small medium grey circles represent hydrogen atoms and the black circles represent carbon ones. $CdCl_4^{2-}$ anions are represented by tetrahedra. Hydrogen bonds are represented by broken lines.

of an infinite one-dimensional inorganic chain of $[Cd_3Cl_{10}]_n^{4n-}$ moieties running along the [011] direction (Figure 8). Two types of six-coordinated Cd are observed: $Cd(1)Cl_6$ and $Cd(2)Cl_6$. In the trimer, two $Cd(2)Cl_6$ octahedra generated by a symmetric center share one bridging chlorine atom (Cl(4), Cl(4')), the $Cd(1)Cl_6$ octahedron shares one bridging chlorine atom (Cl(4), Cl(3)) with the $Cd(2)Cl_6$ octahedron and another bridging chlorine atom (Cl(4), Cl(5)) with

**Table 8: Main inter-atomic distances (Å) and bond angles (°) involved in the hydrogen Bonds of $[C_{12}H_{17}N_2]_2CdCl_4$ compounds. (Esd are given in parentheses).**

| Cl...H-N | H-N (Å) | Cl...H (Å) | Cl...N (Å) | Cl...H-N (°) |
|---|---|---|---|---|
| Cl1...HN21-N21(i) | 0.74(4) | 2.57(4) | 3.306(4) | 171(3) |
| Cl2...HN12-N12 | 0.78(4) | 2.37(4) | 3.134(4) | 167(4) |
| Cl2...HN11-N11 | 0.82(4) | 2.89(4) | 3.691(4) | 163(4) |

- *symmetry code*: i: 1-x, 2-y, 1-z.

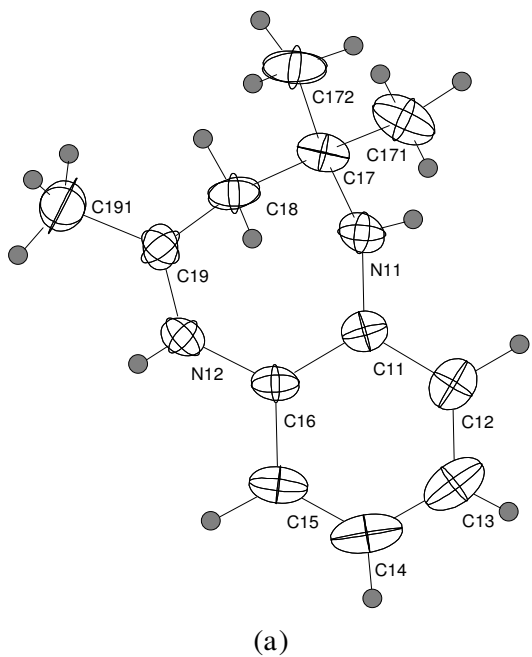

(a)

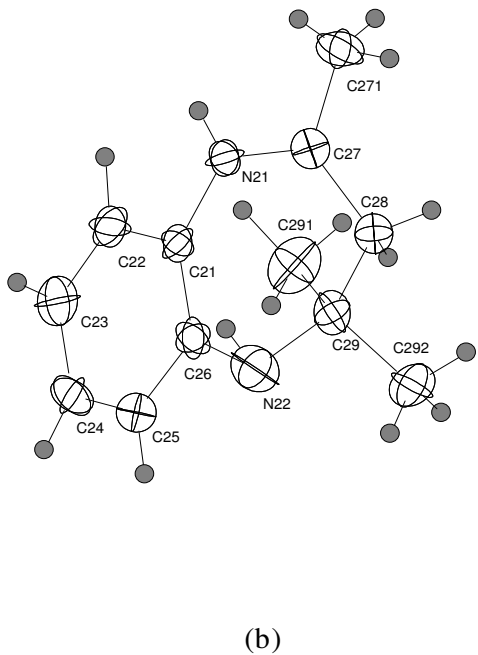

(b)

**Figure 7**
**a:** Showing the ellipsoid of thermic unrest of carbon and nitrogen atoms at 40% in $C_{12}H_{17}N_2^+$(a) cation. **b:** Showing the ellipsoid of thermic unrest of carbon and nitrogen atoms at 40% in $C_{12}H_{17}N_2^+$(b) cation.

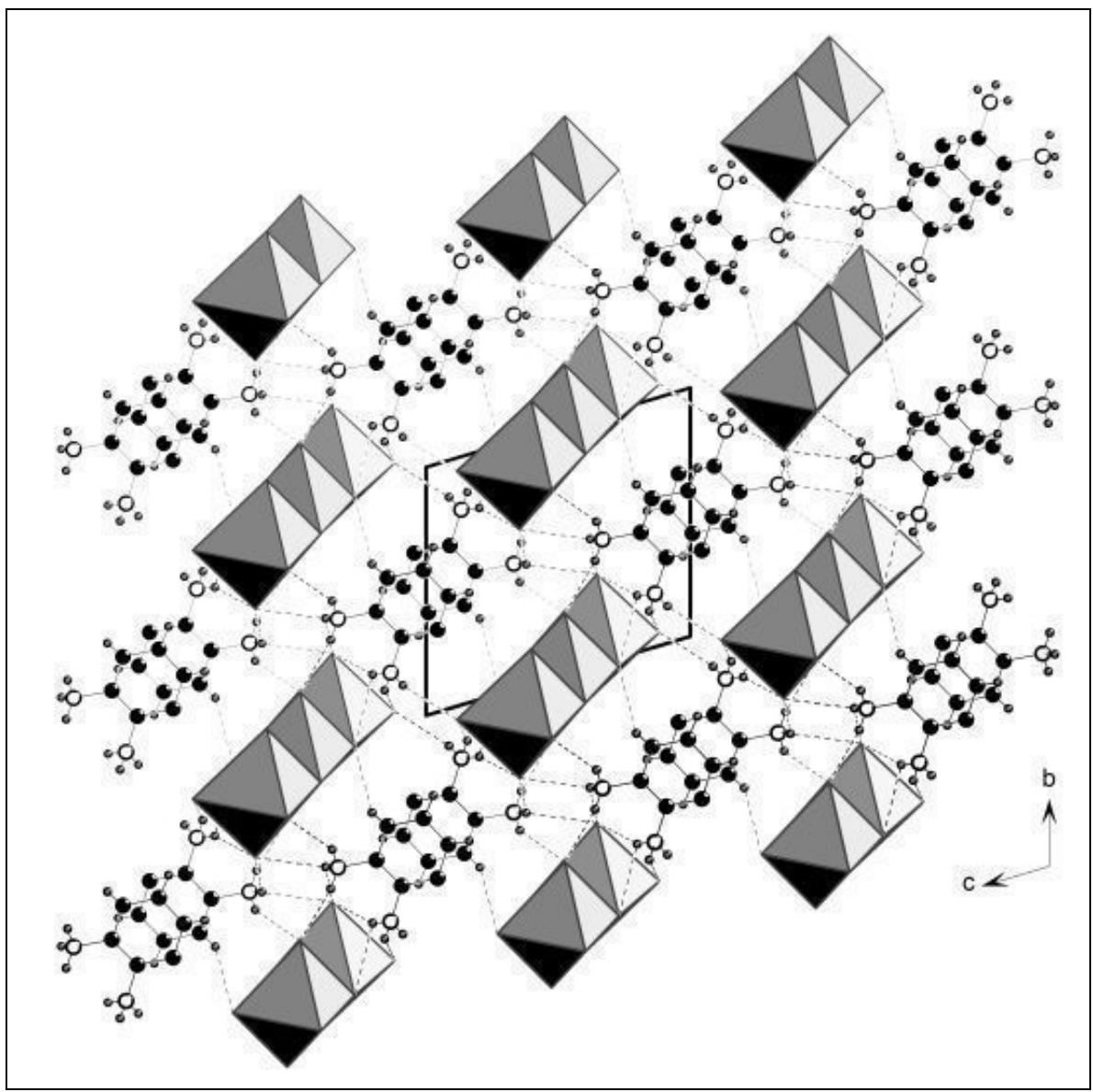

**Figure 8**
[011] view of the structure of $[C_6H_{10}N_2]_2[Cd_3Cl_{10}]$ sample. The large empty circles represent nitrogen atoms, the small medium grey circles represent hydrogen atoms and the black circles represent carbon ones. $Cd_3Cl_{10}^{4-}$ anions are represented by octahedra. Hydrogen bonds are represented by broken lines.

the Cd(2')Cl$_6$ octahedron (Figure 9). In the same layer, the infinite one-dimensional inorganic chains are connected by $C_6H_{10}N_2^{2+}$ cations via hydrogen bonding (N-H...Cl) (Table 9). One type of organic cation, $C_6H_{10}N_2^{2+}$, is observed on both sides of every infinite one-dimensional inorganic chain. Each organic cation orients its NH$_3$ groups towards the inorganic chain in order to form three hydrogen bonds with free chlorine atom of the CdCl$_6$ octahedron and two hydrogen bonds with common chlorine atoms between two types of octahedron.

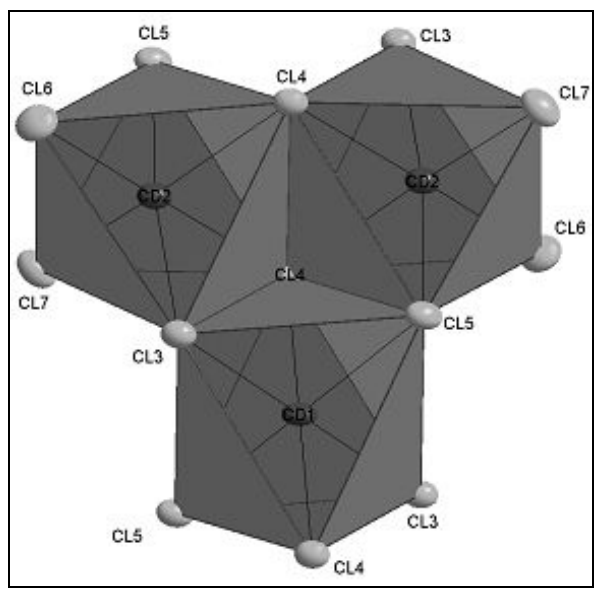

**Figure 9**
Showing the coordination of $Cd_3Cl_{10}^{4-}$ anions and the ellipsoid of thermic unrest of chlorine and cadmium atoms at 40%.

From one organic-inorganic layer to other, the cohesion is assured by hydrogen bonding (N-H...Cl) (table 9). Every cation forms three hydrogen bonds with the two layers observed on both sides of its own layer.

### 3.3.2.1 Geometry and coordination of hexachlorocadmate anion
### 3.3.2.1.1 Geometry of $Cd(1)Cl_6$

The geometry of the $Cd(1)Cl_6^{4-}$ anion is shown in Figure 9, and it is apparent that the coordination around the cadmium is slightly distorted from octahedral symmetry, presenting a $C_{2v}$ punctual symmetry. The base of this octahedron is formed by Cl(3) and Cl(5) chlorine atoms and their symmetry by reversal center. The Cl-Cd-Cl angle values are in the 89.54(4)° – 90.46(4)°

Table 9: Main inter-atomic distances (Å) and bond angles (°) involved in the hydrogen Bonds of $[C_6H_{10}N_2]_2[Cd_3Cl_{10}]$ compounds. (Esd are given in parentheses).

| N-H...Cl | H-N (Å) | Cl...H (Å) | Cl...N (Å) | Cl...H-N (°) |
|---|---|---|---|---|
| N1-HN11... Cl7[i] | 0.923(4) | 2.333(6) | 3.199(8) | 156.3(8) |
| N1-HN12... Cl6 | 1.108(6) | 2.102(7) | 3.122(6) | 151.8(7) |
| N1-HN12... Cl5[i] | 1.108(2) | 3.195(4) | 3.776(8) | 113.5(9) |
| N1-HN13... Cl6[ii] | 0.826(5) | 2.458(4) | 3.185(5) | 147.3(6) |
| N2-HN21... Cl5[iii] | 1.011(4) | 2.191(6) | 3.198(5) | 173.7(7) |
| N2-HN22... Cl5[vii] | 0.751(4) | 2.551(5) | 3.254(7) | 156.5(8) |
| N2-HN22... Cl4 | 0.751(3) | 3.149(4) | 3.548(3) | 116.5(4) |
| N2-HN23... Cl7[i] | 0.992(4) | 2.443(5) | 3.222(4) | 135.1(6) |
| N2-HN23... Cl7[iv] | 0.992(6) | 2.693(6) | 3.310(4) | 120.6(4) |

- <u>symmetry code</u>: i: x+1, y, z; ii: -x+3, -y, -z+2; iii: x+1, y+1, z;
iv: -x+2, -y, -z+1; v: x-1, y-1, z; vi: x-1, y, z; vii: -x+1, -y, -z+1

range (table 6). The range of Cd-Cl distances is between 2.551(1) and 2.669(1)Å (table 4). The chlorine atom observed in both sides of the tetrahedron base show a longer Cd-C1 bond length (Cd-Cl(4) = 2.689(1)Å). The Cl(4)-Cd-Cl angles with the Cl atom of the octahedron base are in the 84.28(3)°–95.72(3)° range, shown to be slightly distorted from octahedral symmetry.

The Cl-Cd-Cl angles of opposite chlorine atoms about cadmium atom center are planar. In fact, the cadmium ion is located at the center of octahedron.

### 3.3.2.1.2 Geometry of Cd(2)Cl$_6$

The Cd(2)Cl$_6$ octahedra presents a C$_i$ punctual symmetry. The Cl-Cd-Cl angles between opposite chlorine atoms about the cadmium atom center do not form the ideal octahedral shape. The angle values are in the 166.97(4)–178.10(4)° range, which proves that the cadmium atom is slightly shifted to the center of the octahedral one (table 6). The remainder of the Cl-Cd-Cl angles show a variation of ± 6° in both sides of ideal octahedral shape (90°). The Cd-Cl distances are more dispersed than those observed in the Cd(1)Cl$_6$ octahedra. The range of Cd-Cl distances is between 2.519(1) and 2.739(1)Å (table 4).

### 3.3.3 Geometry of organic group

The asymmetrical unit contains only one $C_6H_{10}N_2^{2+}$ grouping (Figure 10). The aromatic ring of the cation is slightly distorted. The carbon atoms show a low distortion compared to the average plan of 0.52Å.

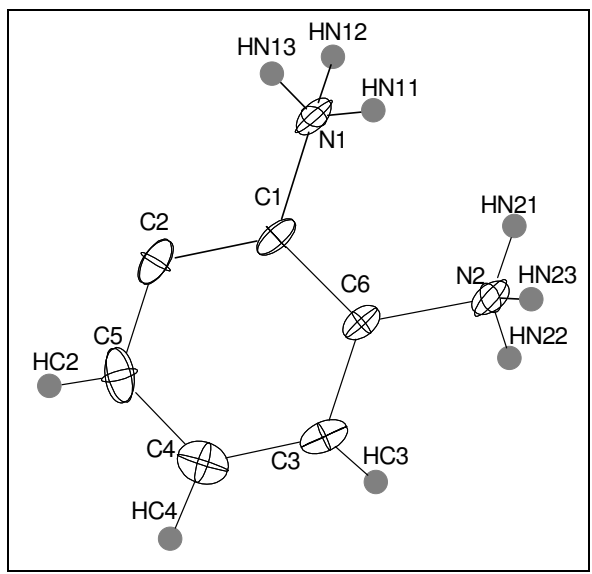

**Figure 10**
Showing the ellipsoid of thermic unrest of carbon and nitrogen atoms at 40% in $C_6H_{10}N_2^{2+}$ cation.

The main geometric features of the organic cation are similar to those commonly observed in hybrid compounds [13]. The C-C distances and the C-C-C angles values in the benzene vary in the range 1.367(7)–1.395(7)Å and 119.1(4)–120.9(5)°, respectively (tables 5 and 7).

## 4. Conclusion

Two new compounds $[C_{12}H_{17}N_2]_2CdCl_4$ and $[C_6H_{10}N_2]_2[Cd_3Cl_{10}]$ have been synthesized using solution methods. The atomic arrangement of the both compounds can be described by alternating layers of organic and inorganic material stacked parallel to the ab plane and according to the [011] direction respectively. The material cohesion for all compounds is assured by two different bonds. The bonding between the inorganic and organic layer is established by N – H...Cl – Cd interaction, and the cohesion of the organic layer is assumed by Van Der Waals interaction between aromatic π-stacking. The inorganic layer for the $[C_6H_{10}N_2]_2[Cd_3Cl_{10}]$ compound is constructed from infinite one-dimensional inorganic chains of $[Cd_3Cl_{10}]_n^{4n-}$ moieties running along the [011] direction, and that for the $[C_{12}H_{17}N_2]_2CdCl_4$ sample is formed from insulated tetrahedrals $[CdCl_4]^{2-}$. Infrared and Raman spectroscopy and NMR study confirms the presence of organic and inorganic groups for both compounds.

**Acknowledgements**
We are grateful to Professor Monsour Salem for informative discussion in order to propose a way to justify obtaining the ion 2,4,4-trimethyl-4,5-dihydro-3H-benzo [b] diazepin-1-ium from 1,2-phenylenediamine, in the synthesis of compound 1.

**References**
1. Kimizuka N, Kunitake T: *Advanced Materials* 1996, **8:**89.
2. Mitzi DB, Chondroudis K, Kagan CR: *IBM Journal of Research and Development* 2001, **45:**1.
3. Barbour LJ, Macgillivray LR, Atwood JL: *Supramolecular Chemistry* 1996, **7:**167.
4. Muller-Dethlefs K, Hobza P: *Chemical Reviews* 2000, **100:**143.
5. Allen FH, Hoy VJ, Howard JAK, Thalladi VR, Desiraju GR, Wilson CC, McIntyre GJ: *Journal of the American Chemical Society* 1997, **119:**3477.
6. Aullon G, Bellamy D, Brammer L, Bruton EA, Orpen AG: *Chemical Communications* 1998:653.
7. Lewis GR, Orpen AG: *Chemical Communications* 1998:1873.
8. Dolling B, Gillon AL, Orpen AG, Starbuck J, Wang X: *Chemical Communications* 2001:567.
9. Battaglia LP, Bonamartini Corradi A, Pelosi G, Cramarossa MR, Manfredini T, Pellacani GC, Motori A, Saccani A, Sandrolini F, Brigatti MF: *Chemistry of Materials* 1992, **4:**813.
10. Veal JT, Hodgson DJ: *Inorganic Chemistry* 1972, **11:**597.
11. Corradi AB, Cramarossa MR, Saladini M: *Inorganica Chimica Acta* 1997, **257:**19.
12. Corradi AB, Ferrari AM, Pellacani GC: *Inorganica Chimica Acta* 1998, **272:**252.
13. Corradi AB, Cramarossa MR, Saladini M, Battaglia LP, Giusti J: *Inorganica Chimica Acta* 1995, **230:**59.
14. Maldonado CR, Quiros M, Salas JM: *Journal of Molecular Structure* 2007 in press.
15. Jian FF, Zhao PS, Wang QX, Li Y: *Inorganica Chimica Acta* 2006, **359:**1473.
16. Thorn A, Willett RD, Twamley B: *Crystal Growth and Design* 2006, **6:**1134.
17. Sheldrick GM: **"SHELXS-86, in crystallographic computing 3".** Edited by: Sheldrick GM, Krüger C, Goddard R. Oxford University Press; 1985.

18. Sheldrick GM: **"SHELXL-93, A Program for the refinement of crystal structures from diffraction data".** *University of Göttingen* 1993.
19. Rai AK, Singh R, Singh KN, Singh VB: *Spectrochimica Acta* 2006, **A63:**483.
20. Marzotto A, Clemente DA, Benetollo F, Valle G: *Polyhedron* 2001, **20:**171.
21. Goggin PL, Goodfellow RJ, Kessler K: *Journal of the Chemical Society, Dalton Transactions* 1977:1914.
22. Mokhlisse R, Couzi M, Lassegues JC: *Journal of Physics C: Solid State physics* 1983, **16:**1353.
23. Ackerman JJH, Orr TV, Bartuska VJ, Maciel GE: *Journal of the American Chemical Society* 1979, **101:**341.
24. Sakida S, Nakata H, Kawamoto Y: *Solid State Communications* 2003, **127:**447.
25. Brown ID: *Acta Crystallographica* 1976, **A32:**24.
26. Luque A, Sertucha J, Castillo O, Roman P: *New Journal of Chemistry* 2001, **25:**1208.
27. Neve F, Francescangeli O, Crispini A: *Inorganica Chimica Acta* 2002, **338:**51.
28. Baur W: *Acta Crystallographica* 1974, **B30:**1195.